\documentclass[prb,twocolumn]{revtex4}% Physical Review B

\usepackage{graphicx}% Include figure files
\usepackage{dcolumn}% Align table columns on decimal point
\usepackage{bm}% bold math
\begin{document}
%\draft

\title{Collective Edge Modes of a Quantum Hall Ferromagnet in Graphene}

\author{V. Mazo$^1$, H. A. Fertig$^2$ and E. Shimshoni$^1$}
\affiliation{Department of Physics, Bar-Ilan University, Ramat Gan 52900, Israel \\
Department of Physics, Indiana University, Bloomington, IN 47405
}
%\author{E. Shimshoni}
%\affiliation{Department of Physics, Bar-Ilan University, Ramat Gan 52900, Israel}
%\author{H.A. Fertig}
%\affiliation{Department of Physics, Indiana University, Bloomington, IN 47405}

\date{\today}

\begin{abstract}
We derive an effective field-theoretical model for the one-dimensional collective mode associated with a domain wall in a quantum Hall ferromagnetic state, as realized in confined graphene systems at zero filling. To this end, we consider the coupling of a quantum spin ladder forming near a kink in the Zeeman field to the spin fluctuations of a neighboring spin polarized two-dimensional environment. It is shown, in particular, that such coupling may induce anisotropy of the exchange coupling in the legs of the ladder. Furthermore, we demonstrate that the resulting  ferromagnetic spin-$1/2$ ladder, subject to a kinked magnetic field, can be mapped to an {\it antiferromagnetic} spin chain at zero magnetic field.
\end{abstract}
\pacs{73.21.-b, 73.22.Gk, 73.43.Lp, 73.22.Pr, 75.10.Pq}

\maketitle

\section{Introduction}\label{Intro}

When an electron is confined within the lowest Landau level in a
quantum Hall (QH) system, its position is described solely by the
guiding center, whose X and Y coordinates do not commute with one
another. Hence, the QH system can be formulated as a dynamical
system in the noncommutative plane \cite{Prange1990,Sarma1997}.
When the system supports discrete degrees of freedom, such as
spin or layer index, for integrally filled Landau levels
quantum coherence develops owing to the exchange
interaction, and the system becomes ferromagnetic. The
characteristic ground state is spin--polarized (or
isospin-polarized, e.g. in bilayer QH systems), and single
spin-flip excitations are not favored due to the large cost in
exchange energy. Instead, the elementary excitation is a
topological soliton named a Skyrmion
\cite{Lee_1990,Sondhi_1993,Fertig_1994,Moon}:
a spin texture, where several
spins are coherently rotated to lower the interaction energy.
Skyrmions are indexed by a quantized topological charge -- the
Pontryagin number of the spin texture, associated also with a
quantized electric charge. Indeed, the experimental detection of
Skyrmions in QH systems realized in semiconductor devices \cite{Barett_1995, Leadley_1997} has
provided compelling evidence for QH ferromagnetism.

In what follows, we study a two component quantum Hall system,
with two Landau levels lying near the Fermi energy, with
enough electrons to fill one of them.
The effective Hamiltonian describing this quantum Hall ferromagnet
(QHFM) is generally of the form
\begin{equation}\label{QHFM}
    H_{\text{QHFM}}= J_s\int d^2r \big[\sum_{a=x,y,z}\big|\nabla
    \mathcal{S}_a (\textbf{r})\big|^2 - \mathcal{B}(\textbf{r})\mathcal{S}_z(\textbf{r})\big];
\end{equation}
here $\vec{\mathcal{S}}$ is the spin field
(with $|\vec{\mathcal{S}}|=1/2$),
and
$J_s$ the spin stiffness determined by the exchange energy. The
local, possibly nonuniform Zeeman field $\mathcal{B}(\textbf{r})$
encodes the noninteracting energy spectrum, which may include
dispersion of the Landau levels due to boundaries or
external potentials. Eq. (\ref{QHFM}) is actually a nonlinear
sigma model with the O(3) symmetry broken down to O(2) symmetry. It
supports a collective spin excitation, which is the Goldstone mode
associated with the spontaneous symmetry breaking. Due to the association of the spin texture with electric charge \cite{Lee_1990}, this mode
may carry charge and can therefore contribute to electric transport
under certain circumstances \cite{Fertig_2006}.
% \begin{equation}
%  Q_{sky}=\frac{1}{8\pi}\int d^2 x \epsilon_{abc}\epsilon_{ij}\mathcal{S}_a\partial_i \mathcal{S}_b \partial_j \mathcal{S}_c,
%\end{equation}

An interesting manifestation of collective states in a QHFM is
expected in graphene\cite{Fertig_2006,Barlas2008,cote08,Zhao2010}. The Dirac dynamics of electrons near the the $\mathbf{K}$ and $\mathbf{K'}$
points in the band structure dictates a unique, particle-hole conjugate spectrum of the Landau levels in the integer quantum Hall regime \cite{CastroNetoRMP}. Most prominently, there exist zero energy Landau level states, responsible for unusual behavior of the $\nu=0$ QH state \cite{Abanin2007,Checkelsky}. In monolayer graphene, the $\nu=0$ state possesses a four-fold degeneracy associated with the
two valleys ($\mathbf{K}$ and $\mathbf{K'}$) and the two spin states. The Zeeman
coupling separates the states into two particle-hole conjugate pairs, above and below zero energy. In bilayer graphene, the layer index degree of freedom of the bilayer system further doubles
the zero energy degeneracy, which can be lifted by applying a perpendicular electric field \cite{McCann2006,McCann2006a}. When interactions are included, the half-filled zero energy
states spontaneously polarize due to exchange, and give rise to a spin or valley polarized ferromagnetic
ground state \cite{Zhang2006,Zhao2012}.

The unusual bulk spectrum of Landau levels in undoped graphene dictates an interesting structure of the edge states near the physical edge of a ribbon \cite{BreyFertig,Abanin_2006,Mazo}, or at the interface between two opposite polarities of the gate voltage in a bilayer system \cite{Paramekanti}. Most prominently, it gives rise to level crossings between an electron-like edge mode with a given spin or isospin state and a hole-like mode with the opposite spin/isospin state, localized on the same edge. This implies a spatial
reversal in the direction of the effective Zeeman field, which in the presence of interactions induces a coherent domain wall (DW) between regions with distinct configurations of the QHFM ground state \cite{Fertig_2006,Huang}. The resulting QHFM state is a realization of the model Eq. (\ref{QHFM}) with $\mathcal{B}(\textbf{r})=\mathcal{B}(x)$, in which $\mathcal{B}(x)$ changes sign across a line in the $xy$ plane. In this geometry, quantum fluctuations of the spin/isospin rotation angle support a
collective edge mode, which possesses a one-dimensional (1D) dynamics.
This edge mode has been argued to behave at low energies as a Luttinger liquid \cite{Fertig_2006,bilayerLL}, or alternatively as an anti-ferromagnetic (AFM) spin chain \cite{Shimshoni2009}. However, an explicit derivation of the 1D effective model from the two-dimensional (2D) QHFM [Eq. (\ref{QHFM})] has not been carried out in earlier literature beyond the semiclassical spin-wave approximation.

In the present paper, we consider a simple model for a 2D QHFM subject to a kink in the Zeeman field, which allows the derivation of an effective 1D quantum field-theoretical model for the dynamics of the collective DW mode along the kink. We find that within an appropriate regime of parameters, in particular assuming a sufficiently strong Zeeman field in the polarized regions, the low energy dynamics is equivalent to an AFM spin-$1/2$ chain, whose parameters can be systematically related to the original 2D system.

This paper is organized as follows: in Section \ref{Derivation}, we study the coupling of a single quantum spin ladder forming near a kink in the Zeeman field to the spin fluctuations of a neighboring spin polarized 2D environment. It is shown that the resulting effective 1D theory manifests anisotropy of the exchange coupling in the legs of the ladder. In Section \ref{AFMSpinChain}, we consider a ferromagnetic spin-$1/2$ zigzag ladder subject to a staggered magnetic field, and demonstrate its mapping to an {\it antiferromagnetic} spin chain at zero magnetic field. Finally, some concluding remarks are presented in Section \ref{Summary}.

\section{Derivation of a Quasi 1D Model for a Domain Wall}\label{Derivation}

\begin{figure}[t!]
\vspace{-10pt}
\begin{center}
\scalebox{0.3}{{\includegraphics{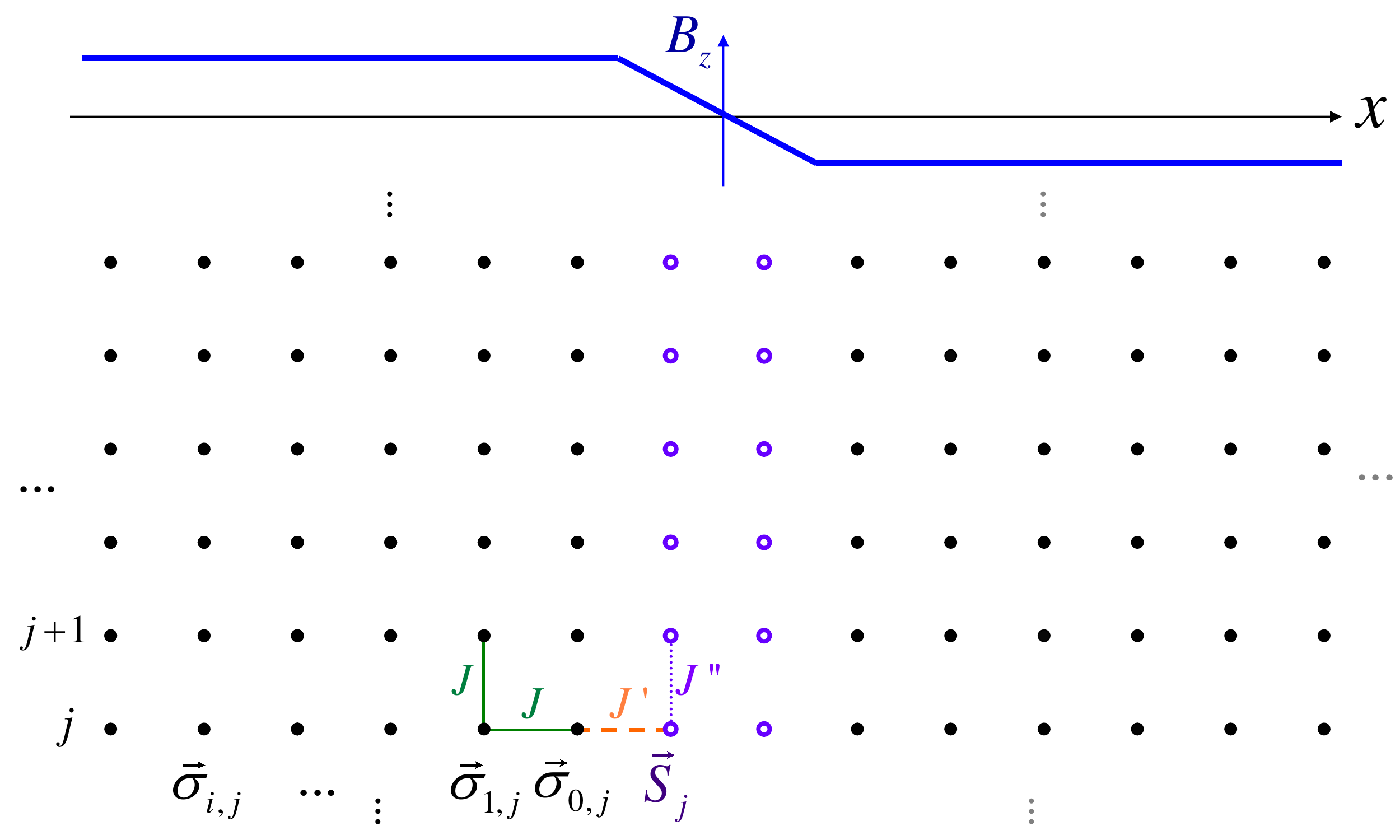}}}\caption{A symplified two dimensional system that consists of spins $\vec{\sigma}_{i,j}$ at high magnetic field and spins $\vec{S_j}$ at lower magnetic field.
}\label{2D_spinsystem}
\end{center}
\vspace{-10pt}
\end{figure}

We consider a 2D electron system in a QHFM state, described by a discrete version of Eq. (\ref{QHFM}) where the lattice spacing between local spin operators is set by the average distance between electrons, proportional to the magnetic length $\ell_B=\sqrt{\hbar c/eB}$. The magnetic field $B_z({\bf r}=(x,y))$ is assumed to be independent of $y$, and
to change sign across a narrow strip near $x=0$ as depicted in Fig. \ref{2D_spinsystem}.
The mean-field groundstate of this system contains an in-plane component to
$\vec{\mathcal{S}}({\bf r})$ near $x=0$, so that the O(2) symmetry of the
Hamiltonian is broken, and a gapless one-dimensional collective mode, propagating
along the $y$-direction, is present.  While quantum fluctuations restore the
broken symmetry, the quasi-one-dimensional mode remains in the spectrum.
Its dynamics, however, is affected by the ferromagnetic coupling to the bulk spins, composed of two semi-infinite planes each subject to a uniform magnetic field. Quantum fluctuations of these bulk spins act as an environment. Below we integrate over these degrees of freedom, to derive an effective model for the quasi 1D interface spin degrees of freedom.

For simplicity we focus on the square lattice model depicted in Fig. \ref{2D_spinsystem}.
Local spin $\frac{1}{2}$ operators at $|x|>x_0$, denoted $\sigma_{i,j}$, are subject to a constant magnetic field $B$, and spins at $-x_0<x<x_0$, denoted $\vec{S}_{i,j}$, are subject to a nonuniform magnetic field which changes linearly from $B$ to $-B$. We will consider only the left semi-infinite plane ($x<0$) and assume that $x_0\sim\ell_B$, so that the region $-x_0<x<0$ contains only a single chain of spins $\vec{S_j}$, subject to a uniform magnetic field $B_{1D}<<B$. The corresponding Hamiltonian is
\begin{eqnarray}\label{Hamitonian}
H &=& H_{env}+H_{1D}+J' \sum\limits_{j=-\infty}^{\infty}\vec{\sigma}_{0,j}\cdot\vec{S}_{j}\; ,\\ \nonumber
H_{env} &=& J \sum\limits_{j=-\infty}^{\infty}\sum\limits_{i=0}^{\infty}\big[\vec{\sigma}_{i+1,j}\cdot\vec{\sigma}_{i,j}
+\vec{\sigma}_{i,j+1}\cdot\vec{\sigma}_{i,j}\big] \\
&-& B\sum\limits_{j=-\infty}^{\infty}\sum\limits_{i=0}^{\infty}{\sigma}_{i,j}^z \\
 H_{1D} &=& J'' \sum\limits_{j=-\infty}^{\infty}\vec{S}_{j}\cdot\vec{S}_{j+1}
- B_{1D}\sum\limits_{j=-\infty}^{\infty}{S}_{j}^z
\end{eqnarray}
where all couplings are ferromagnetic ($J, J', J'' < 0$).  (Note the labeling scheme
used for $\sigma_{i,j}$ as depicted in Fig. \ref{2D_spinsystem}.)

Assuming the bulk magnetic field $B$ to be very high, a spin wave approximation can be used for the environmental spins $\vec{\sigma}_{i,j}$ \cite{assa}:
\begin{equation}
    \sigma_{i,j}^z\approx s_0-\frac{\left(\sigma_{i,j}^x\right)^2+\left(\sigma_{i,j}^y\right)^2}{2s_0}
\end{equation}
where the total spin $s_0$ (with actual value $s_0=1/2$) is maintained as a parameter, playing the role of $\hbar$ in the canonical quantization of the spin fields in the $xy$-plane, which obey  $[\sigma_{i,j}^x,\sigma_{i,j}^y]=is_0$ in the spin-wave
approximation.  This yields the quadratic action for the isolated semi-infinite spin environment
\begin{eqnarray}\label{Action}
& S_{env}[\vec{\sigma}]=\int\limits_0^{\beta}d\tau\sum\limits_{j=-\infty}^{\infty}\sum\limits_{i=0}^{\infty}\, \Big\{\frac{i}{s_0} \sigma^{x}_{i,j}\partial_{\tau}\sigma_{i,j}^{y}
+J\Big({\sigma}_{i+1,j}^{x}{\sigma}_{i,j}^x \nonumber \\ \nonumber
& \qquad
+{\sigma}_{i+1,j}^{y}{\sigma}_{i,j}^{y}
+{\sigma}_{i,j+1}^{x}{\sigma}_{i,j}^x
+{\sigma}_{i,j+1}^{y}{\sigma}_{i,j}^y\Big)\\ \nonumber
& \quad \quad
+\Big(\frac{1}{2s_0}B-2J\Big)\big(({\sigma}_{i,j}^x)^2 +({\sigma}_{i,j}^y)^2\big) \Big\}\\
& \quad \quad
+\frac{J}{2} \int\limits_0^{\beta}d\tau\sum\limits_{j=-\infty}^{\infty}\,\big(({\sigma}_{0,j}^x)^2 +({\sigma}_{0,j}^y)^2\big)
\end{eqnarray}
where $\vec{\sigma}=(\sigma^x, \sigma^y)$, and $\beta$ is the inverse temperature.

We next notice that the 1D spin chain $\vec{S}$ is coupled to the environment via the single chain $\vec{\sigma}_0$, whose effective 1D action can be expressed as
\begin{equation}\label{S_env1D}
S_{env}^{1D}[\vec{\sigma}_0]=S^{SW}[\vec{\sigma}_0]+\delta S[\vec{\sigma}_0]\; ,
 \end{equation}
where
\begin{eqnarray}\label{S_SW}
& S^{SW}[\vec{\sigma}_0]
=\int\limits_0^{\beta}d\tau\sum\limits_{j}\big[\frac{i}{s_0}\sigma_{0,j}^x\partial_{\tau}\sigma_{0,j}^y+J\vec{\sigma}_{0,j}\cdot\vec{\sigma}_{0,j+1} \nonumber \\
& \qquad +\frac{1}{2s_0}B\big(({\sigma}_{0,j}^x)^2 +({\sigma}_{0,j}^y)^2\big) \big]
\end{eqnarray}
and $\delta S$ is obtained after trace over the remaining environmental spins $\vec{\sigma}_{i\geq1}$:
\begin{equation}
e^{-\delta S[\vec{\sigma}_0]}=\int D\vec{\sigma}_{i\geq1} \, e^{-S_{env}[i\geq 1]}e^{-S_{env}[0,1]}\; .
\end{equation}
Here $S_{env}[i\geq 1]$ schematically denotes the part of $S_{env}[\vec{\sigma}]$ describing the spins $\vec{\sigma}_{i,j}$ with $i\geq1$, and $S_{env}[0,1]$ contains the interactions between spins $\vec{\sigma}_{0,j}$ and $\vec{\sigma}_{1,j}$.

To carry out the integration, we wish to use a Fourier representation of the spin wave fields. Since the semi-infinite plane imposes
inconvenient boundary conditions, we employ a duplication of the chains $i\geq1$ via the relation
\begin{eqnarray}\label{double_env}
& e^{-2\delta S[\vec{\sigma}_0]}=e^{S^{SW}[\vec{\sigma}_0]}\int D\vec{\sigma}_{i\not= 0}\, %\delta\big(\vec{\sigma}_{0,j}(\tau)-\vec{\sigma}_{0,j}^{(0)}(\tau)\big)
e^{-S_{2D}[\vec{\sigma}]}
\end{eqnarray}
where $S_{2D}[\vec{\sigma}]$ describes the spin-wave action of a full 2D lattice, in the presence of a constant Zeeman field $B$. The spins $\vec{\sigma}_{0}$
are excluded from the integration in Eq. (\ref{double_env}). To enforce this constraint, we introduce Lagrange multipliers in terms of an auxiliary field $\vec{\lambda}$, yielding
\begin{eqnarray}\label{PartFunction}
& e^{-2\delta S[\vec{\sigma}_0]}= e^{S^{SW}[\vec{\sigma}_0]}\int D\vec{\lambda} \, e^{-i\sum\limits_{{k_y},\omega_n}\vec{\lambda}^T({-k_y},-\omega_n)\vec{\sigma}_0({k_y},\omega_n)} \nonumber \\
&\quad \times\int D\vec{\sigma}\,
e^{i\sum\limits_{{k}_y,\omega_n}\vec{\lambda}^T({-k}_y,-\omega_n)\vec{\sigma}({k}_y,\omega_n)
-S_{2D}[\vec{\sigma} ]}
%\\ \nonumber
%& \quad=e^{S^{SW}[\vec{\sigma}_0]-\delta S_{eff}[\vec{\sigma}_0]}
\end{eqnarray}
where we have used the Fourier transforms
\begin{eqnarray}
\vec{\sigma}_{i,j}(\tau)
&=&\frac{1}{\sqrt{N_xN_y\beta}}\sum_{\vec{k},\omega_n}e^{i(\vec{k}\cdot\vec{R}_{i,j}-\omega_n\tau)}
   \vec{\sigma}(\vec{k},\omega_n)\; , \nonumber \\ \vec{R}_{i,j}&=& a(i\,\hat{x}+j\,\hat{y})\; , \nonumber \\
\vec{\lambda}_{j}(\tau)
&=&\frac{1}{\sqrt{N_y\beta}}\sum_{k_y,\omega_n}e^{i(k_yR_j-\omega_n\tau)}\vec{\lambda}(k_y,\omega_n)
\end{eqnarray}
(with $N_x$, $N_y$ the total number of sites in the corresponding directions). The bulk action in Eq. (\ref{PartFunction}) can be expressed in terms of these fields as
%\begin{widetext}
\begin{equation}\label{S_2D}
S_{2D}[\vec{\sigma}]=\frac{1}{2s_0}\sum_{\vec{k},\omega_n}\vec{\sigma}^{T}(-\vec{k},-\omega_n)
   G_{2D}^{-1}(\vec{k},\omega_n)\vec{\sigma}(\vec{k},\omega_n)
\end{equation}
%\end{widetext}
with
\begin{eqnarray}\label{G_2Dinv}
 G_{2D}^{-1}(\vec{k},\omega_n) &=&
 \left( \begin{array}{cc}
 \omega_{2D}(\vec{k}) & -\omega_n \\
  \omega_n  &  \omega_{2D}(\vec{k})
\end{array}\right)  \; ,
\\ \nonumber
  \omega_{2D}(\vec{k}) &=&
 2Js_0\big[\cos(k_x)+\cos(k_y)-2 \big]+B \\ \nonumber
& \cong & |J|s_0|\vec{k}|^2+B,
\end{eqnarray}
where $\vec{k}$ is in units of $1/a$, and in the last step we have used the long wavelength approximation $|\vec{k}|\ll 1$.
After integration (see Appendix A for details) and substitution in Eq. (\ref{S_env1D}), we obtain
\begin{widetext}
\begin{eqnarray}\label{G_env1D}
& \delta S_{env}^{1D}[\vec{\sigma}_0]
=\frac{1}{2s_0}\sum\limits_{{k_y},i\omega_n}\vec{\sigma}_0(-{k_y},-\omega_n)^T
   (G_{env}^{1D}({k_y},\omega_n))^{-1}\vec{\sigma}_0({k_y},\omega_n)\; ,\\ \nonumber
& (G_{env}^{1D}({k_y},\omega_n))^{-1}\equiv\frac{1}{2}\left( \begin{array}{cc}
-i|J|s_0(k_-+k_+)+\omega_{sw}(k_y) &  |J|s_0(k_+-k_-)-\omega_n    \\ \\
-|J|s_0(k_+-k_-)+\omega_n  &  -i|J|s_0(k_-+k_+)+\omega_{sw}(k_y) \\ \\
\end{array}\right)
\end{eqnarray}
where
\begin{eqnarray}
k_{\pm}&=& i\sqrt{k_y^2+\frac{(B\pm i\omega_n)}{|J|s_0}}
\cong i\sqrt{\frac{B}{|J|s_0}}+\frac{i}{2}\sqrt{\frac{|J|s_0}{B}}k_y^2\mp\frac{\omega_n}{2\sqrt{B|J|s_0}} \; ,\\ \nonumber
\omega_{sw}(k_y)&=& 2Js_0[\cos(k_y)-1]+B\cong |J|s_0 k_y^2+B\; .
\end{eqnarray}
\end{widetext}
Inserting the last approximations in Eq. (\ref{G_env1D}), we note that the resulting effective action of the spin chain $\vec{\sigma}_{0}$ has the form of a semiclassical spin-wave theory in 1D with renormalized parameters:
\begin{eqnarray}\label{alpha_def}
\tilde{s}_0&=&s_0\left(\frac{2}{1+\alpha}\right)\; ,\quad\tilde{J}=J\left(\frac{1+\alpha}{2}\right)\; ,
\nonumber \\ \tilde{B}&=& B\left(\frac{1+2\alpha}{1+\alpha}\right)\; ,\quad{\rm with}\quad\alpha \equiv \sqrt{\frac{|J|s_0}{B}}\; .
\end{eqnarray}
The fractional renormalization of the spin magnitude $s_0$ is a signature for the deviation from a pure spin Hamiltonian dynamics, arising from the trace over environmental degrees of freedom.

We are finally ready to derive the effective action for the spin chain $\vec{{S}}$, obtained after integration over the spins $\vec{\sigma}_0$:
%\begin{widetext}
\begin{eqnarray}\label{S_eff}
e^{-S_{eff}[\vec{S}]}&=&e^{-S_0[\vec{S}]-\delta S_{eff}[\vec{S}]} \\ \nonumber
&=&e^{-S_0[\vec{S}]} \int D\vec{\sigma}_0 \, e^{-S_{int}[\vec{\sigma}_0,\vec{S}]-S_{env}^{1D}[\vec{\sigma}_0]}\; ,
\end{eqnarray}
%\end{widetext}
in which $S_{int}[\vec{\sigma}_0,\vec{S}]$ describes the coupling between the two chains, associated with the last term in Eq. (\ref{Hamitonian}).
Since we wish to account for the full quantum mechanical nature of the spin operators $\vec{{S}}$, a spin-wave approximation of the latter is avoided. Hence, a convenient representation of $S_{eff}[\vec{S}]$ in Fourier space is not available. To facilitate a coherent states path-integral formulation, we therefore map the spin operators to interacting fermions via the Jordan-Wigner (JW) transformation \cite{giamarchi}
\begin{eqnarray}
 & S_j^+=c_j^{\dagger}e^{i\pi\sum\limits_{i<j}c_i^{\dagger}c_i}\; ,\; S_j^-=e^{-i\pi\sum\limits_{i<j}c_i^{\dagger}c_i}c_j
 \nonumber \\
 & S_j^z=c_j^{\dagger}c_j-\frac{1}{2}\; .
\label{JWtransformation}
\end{eqnarray}
Within a spin-wave approximation for the spin fields $\vec{\sigma}_0$, the interaction Hamiltonian acquires the form
\begin{widetext}
\begin{equation}\label{H_int}
 H_{int}[\vec{\sigma_0}, \vec{S}]=J'\sum\limits_{j} \left\{\frac{1}{2} (\sigma_0^+ e^{-i\pi\sum\limits_{i<j}c_i^{\dagger}c_i}c_j + c_j^{\dagger} e^{i\pi\sum\limits_{i<j}c_i^{\dagger}c_i}\sigma_0^-)+s_0(c_j^{\dagger}c_j-\frac{1}{2})\right\}
\end{equation}
\end{widetext}
where $\sigma^{\pm}_0=(\sigma_0^x\pm i\sigma_0^y)$. The last term describes a simple shift of the Zeeman field (i.e., a chemical potential of the JW fermions).  However the coupling of the spin wave fields $\sigma^{\pm}_0$ to the JW fermions is highly non-local and non-linear. Introducing
the variables
 \begin{equation}
\bar{\mathcal{S}}(j,\tau)=\bar{\psi}_j e^{i\pi\sum\limits_{i<j}\bar{\psi}_i\psi_i}
\end{equation}
and its complex conjugate $\mathcal{S}$, which represent the spins in terms of the Grassmann variables $\psi_j$, $\bar{\psi}_j$, and
performing the integration in Eq. (\ref{S_eff}), we find the correction to the action of 1D chain of spins $\vec{S}$ (see Appendix B for details):
\begin{widetext}
\begin{eqnarray}
&  \delta S_{eff}[\vec{S}]=\int d\tau d\tau' \sum\limits_{j,j'}\, \left\{\bar{\mathcal{S}}(j,\tau)V_{eff}(j,j';\tau,\tau')
\mathcal{S}(j',\tau')+c.c.\right\}+J'\int d\tau\sum\limits_j\,(\bar{\psi}_j(\tau)\psi_j(\tau)-\frac{1}{2})\; , \label{deltaSeff} \\
&  V_{eff}(j,j';\tau,\tau')
=-\frac{J'^2\tilde{s_0}}{8}\frac{1}{\sqrt{\pi|J|s_0(\tau-\tau')}}
e^{-\tilde{B}(\tau-\tau')}
 e^{-\frac{(j-j')^2}{4|J|s_0(\tau-\tau')}}\Theta(\tau-\tau') \label{Veff}
\end{eqnarray}
\end{widetext}
where the parameters $\tilde{s_0}$ and $\tilde{B}$ are defined in Eq. (\ref{alpha_def}).

The effective interaction term in Eq. (\ref{Veff}) appears to be hardly useful in its exact form. However, it should be noticed that $V_{eff}$ decays exponentially for $(\tau-\tau')>1/\tilde{B}$. As long as one is interested in physical properties (e.g., spin-spin correlations) in the long length scale limit (or, equivalently, for low temperatures $T\ll \tilde{B}$), $V_{eff}$ may be treated as almost local in imaginary time. In addition, it is short-range in space: the Gaussian factor decays on length scales
\begin{equation}
\tilde{a}\sim a\sqrt{|J|s_0(\tau-\tau')}\sim \alpha a\nonumber
\end{equation}
[$\alpha$ defined in Eq. (\ref{alpha_def})],
i.e. the short distance cutoff is normalized by a constant factor.
%Indeed, a perturbative study of the induced correction to the spin-spin correlation function
%(see Appendix C for details) indicates that the effective interaction is a marginal operator,
%giving rise to a fixed normalization of the power-law which governs their decay.
For $J'<\tilde{B}$, one obtains a converging perturbation series which indicates that $\delta S_{eff}$ is a marginal operator under renormalization group (RG). Its contribution therefore amounts to additive corrections to parameters of the standard terms in the bare action of the quantum spin chain, $S_0$.

The most obvious correction induced by $\delta S_{eff}$ is the modification of the Zeeman field due to the mean--field polarization of the environmental spins: $B_{1D}\rightarrow B_{1D}+|J'|s_0$. More interestingly, the exchange coupling in the $xy$-plane is modified: $J''_{xy}\rightarrow J''_{xy}+\delta J_{xy}$, where
\begin{equation}
\delta J_{xy}\sim -\frac{J'^2}{\tilde{B}}\frac{\tilde{s_0}}{4}=
-\frac{J'^2}{\sqrt{B}(2\sqrt{|J|s_0}+\sqrt{B})}\; .
\end{equation}
Since $J''_z$ is unchanged, this implies that anisotropy is induced in the $xy$-plane. As the bare Heisenberg exchange is ferromagnetic ($J''<0$), the negative correction $\delta J_{xy}$ leads to enhancement of $J''_{xy}$ compared to $J''_z$. As a result, the effective low-energy model for the spin chain is the XXZ model, in the regime where its dynamics is governed by a Luttinger liquid Hamiltonian with a finite Luttinger parameter $K>1$. As discussed in the next Section, this enables the application of Bosonization for the study of its quantum dynamics when coupled to a second chain on the right hand side of $x=0$ (see Fig. 1).

%______________________________________________________________________________________________________

\section{Mapping to AFM Spin Chain}\label{AFMSpinChain}

In the previous section we considered only half of the space, and by integrating
out reservoir degrees of freedom we arrived at a one dimensional spin chain with renormalized couplings. A similar procedure applied to the second half-space yields a parallel spin-chain with identical exchange parameters, but an opposite sign of the effective magnetic field. The two chains are coupled via ferromagnetic exchange interactions. We thus obtain an effective ferromagnetic spin-$1/2$ ladder, subject to a magnetic field which has opposite sign on the two legs, and therefore tends to frustrate the ferromagnetic interactions.

\begin{figure}[t!]
\vspace{-10pt}
\begin{center}
\scalebox{0.3}{{\includegraphics{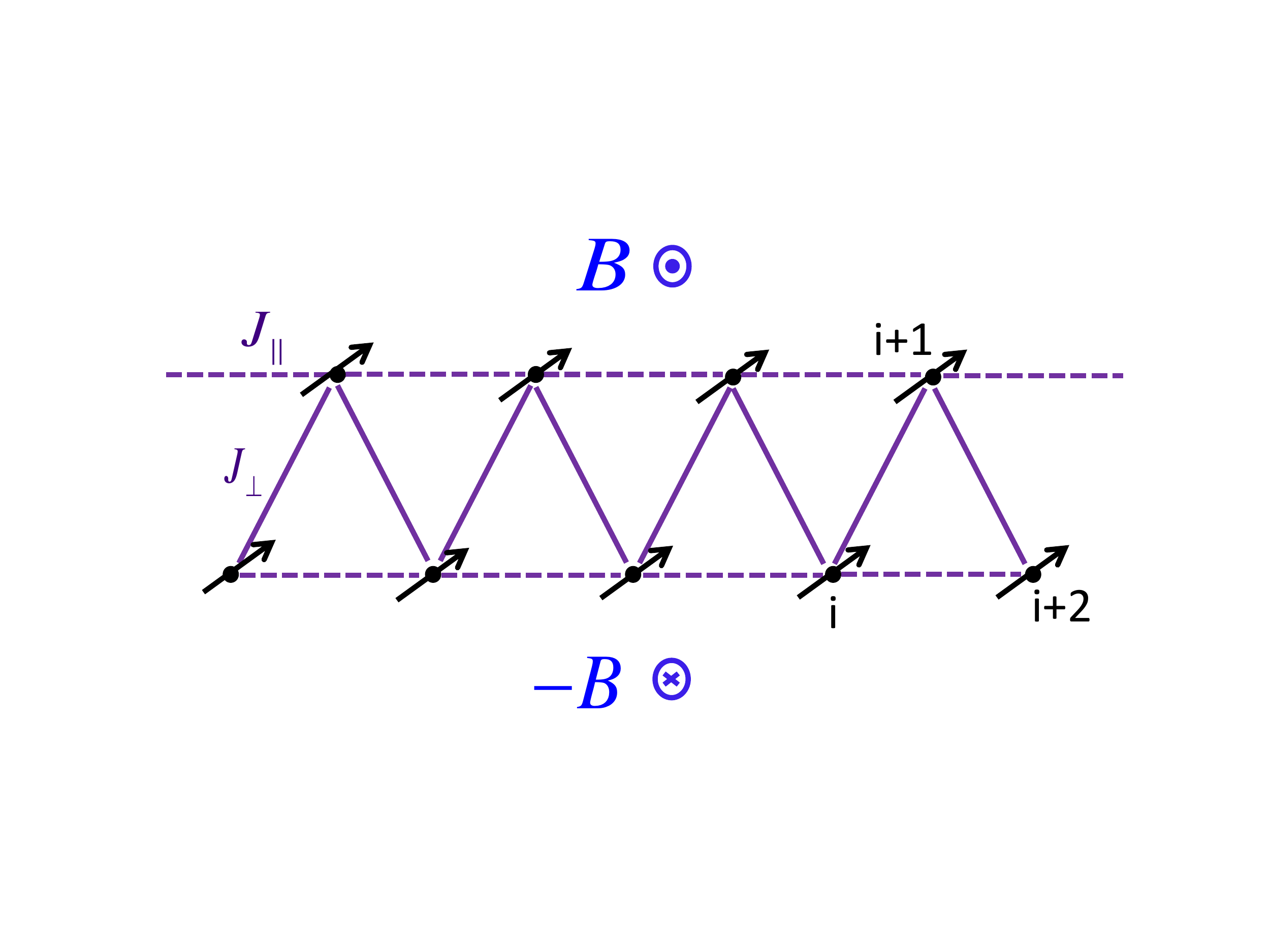}}}\caption{
Ferromagnetic zigzag spin ladder subject to a staggered magnetic field $B$. Transverse exchange coupling ($J_{\perp}$) and longitudinal exchange coupling ($J_{\parallel}$) are represented by full and dashed bonds, respectively.}\label{zigzag}
\end{center}
\vspace{-10pt}
\end{figure}

To study the dynamics of such a system, we focus on the simplest version of a ladder which possesses a zigzag structure (see Fig. 2). To this end, we consider the following model:
\begin{eqnarray}\label{spinchain} \nonumber
& H=H_{\perp}+H_{\parallel}+H_{mag}\; , \\ \nonumber
& H_{\perp} =-\frac{1}{2}|J_{\perp}^{xy}|\sum\limits_{i}\big[S^+_{i}S^-_{i+1}+S^-_{i}S^+_{i+1} \big]
-|J_{\perp}^z|\sum\limits_{i}S^z_{i}S^z_{i+1}\\ \nonumber
& H_{\parallel}= -\frac{1}{2}|J_{\parallel}^{xy}|\sum\limits_{i}\big[S^+_{i}S^-_{i+2}+S^-_{i}S^+_{i+2} \big]
-|J_{\parallel}^z|\sum\limits_{i}S^z_{i}S^z_{i+2}\\
& H_{mag}=- B\sum\limits_i (-1)^i S_i^z\; ,
\end{eqnarray}
where even and odd sites $i$ are located on the top and bottom legs of the ladder, respectively. The zigzag ladder is thus represented as a single chain with nearest and next nearest neighbor interactions, subject to a
staggered magnetic field $B$.
Using the Jordan-Wigner transformation [Eq. (\ref{JWtransformation})] and a subsequent bosonization of the fermion fields in the continuum limit \cite{giamarchi}
\begin{eqnarray}\label{FtoB} \nonumber
& \psi(x)=\psi_R+\psi_L  \\
& {\rm where}\quad \psi _{R,L} = \frac{1}{{\sqrt {2\pi a } }}e^{\pm ik_Fx}e^{i( \mp \phi  + \theta )}
\end{eqnarray}
(in which $\phi$, $\theta$ obey the canonical commutation relations $[\phi(x),\partial_x\theta(x^\prime)]=i\pi\delta(x^\prime-x)$, and $k_F=\pi/2a$), we obtain
\begin{eqnarray}\label{bosonzigzag}\nonumber
& H_{\perp}=\frac{1}{2\pi}\int dx\,\Big[u_{\perp}K_{\perp}(\nabla\theta)^2+\frac{u_{\perp}}{K_{\perp}}(\nabla\phi)^2\Big]\\ \nonumber
& \qquad
+\frac{|J_{\perp}^z|}{2\pi^2 a}\int dx\, \cos(4\phi)\\ \nonumber
& H_{\parallel}= \frac{1}{2\pi}\int dx\Big[u_{\parallel}K_{\parallel}(\nabla\theta)^2\\ \nonumber
& \qquad
+\frac{u_{\parallel}}{K_{\parallel}}(\nabla\phi)^2\Big]+\frac{|J_{\parallel}^z|}{2\pi^2 a}\int dx \,\cos(4\phi)\\ \nonumber
& H_{mag}= - \frac{B}{\pi a}\int dx\, \cos(2\phi)\; .
\end{eqnarray}
Here the velocities $u_{\perp}$, $u_{\parallel}$ and the Luttinger parameters $K_{\perp}$, $K_{\parallel}$ are dictated by the values of $J_{\perp}^\alpha$, $J_{\perp}^\alpha$. For $|J^z|\ll|J^{xy}|$,
\begin{eqnarray}\label{uK}
& u_{\perp}K_{\perp}=v_F^{\perp}=|J^{xy}_{\perp}|a \\ \nonumber
& \qquad \frac{u_{\perp}}{K_{\perp}}=v_F^{\perp}-\frac{4|J^{z}_{\perp}|a}{\pi}\\ \nonumber
& u_{\parallel}K_{\parallel}= \frac{u_{\parallel}}{K_{\parallel}}=0\; .
\end{eqnarray}
More generally, the Hamiltonian can be written as
\begin{eqnarray}\label{Hamiltonian}
& H=\frac{1}{2\pi}\int dx\,\Big[uK(\nabla\theta)^2+\frac{u}{K}(\nabla\phi)^2\Big]
 \\ \nonumber
& \qquad +\frac{|J_{\perp}^z|+|J_{\parallel}^z|}{2\pi^2 a}\int dx\, \cos(4\phi)- \frac{B }{\pi a}\int dx\, \cos(2\phi)
\end{eqnarray}
where
\begin{eqnarray}
& uK = u_{\perp}K_{\perp}+ u_{\parallel}K_{\parallel} \qquad \quad \frac{u}{K}=\frac{u_{\perp}}{K_{\perp}}+\frac{u_{\parallel}}{K_{\parallel}}\; .
\end{eqnarray}
The first term is a Luttinger liquid, with a Luttinger parameter given by
\begin{eqnarray}
& K=\sqrt{\frac{u_{\perp}K_{\perp}+u_{\parallel}K_{\parallel}}{\frac{u_{\perp}}{K_{\perp}}
+\frac{u_{\parallel}}{K_{\parallel}}}}\; .
\end{eqnarray}
Note that in any case $K>1$, characteristic of a ferromagnetic XXZ spin-chain.

We now consider the effect of the non-linear terms in Eq. (\ref{Hamiltonian}). Since the scaling dimension of an operator of the form $\cos(n2\phi)$ is $\Delta_n=n^2K$ (see, e.g., Ref. \onlinecite{giamarchi}), the $\cos(4\phi)$ term is less relevant and can be ignored. The Hamiltonian therefore reduces to a sine-Gordon model,
where the $\cos(2\phi)$ induced by the staggered field becomes relevant when $K$ is tuned below $K_c=2$.

It is interesting to note that the model can be mapped to the continuum limit of an antiferromagnetic (AFM) spin-chain model by rescaling the fields $\phi$, $\theta$ and the parameter $K$ as follows:
\begin{eqnarray}
& \tilde{\phi}=\frac{1}{2}\phi   \qquad \qquad \tilde{\theta}=2\theta   \qquad \qquad   \tilde{K}=\frac{1}{4}K\; .
\end{eqnarray}
In terms of the new fields,
\begin{eqnarray}\label{AFM}
& H=\frac{1}{2\pi}\int dx\,\Big[u\tilde{K}(\nabla\tilde{\theta})^2+\frac{u}{\tilde{K}}(\nabla\tilde{\phi})^2\Big]\\ \nonumber
& \qquad -g\int dx\, \cos(4\tilde{\phi})
\end{eqnarray}
where $g=\frac{B}{\pi a}$.
For $K<4$, one obtains $\tilde{K}<1$ and the model Eq. (\ref{AFM}) can be interpreted as an effective {\it antiferromagnetic} XXZ spin-chain. Tuning the original Luttinger parameter below $K_c=2$ corresponds to $\tilde{K}<1/2$, where the cosine term becomes relevant and a spin gap is opened, as proposed in Ref. \onlinecite{Shimshoni2009}. In the ordered (gapped) phase, $S_z$ is polarized by the staggered field forming a staggered pattern on the zigzag chain, which indeed is equivalent to AFM ordering.

%______________________________________________________________________________________________________

\section{Summary}\label{Summary}

The unique spectral properties of electrons in undoped graphene provide a possibility to realize and control spin-textures and domain walls, forming near a kink-like structure in the effective magnetic field. Quantum fluctuations of the spin configuration dictated by such a kink are manifested by the presence of an effectively 1D collective mode, which propagates along the DW (i.e., in the translationally invariant direction). Its quantum dynamics is governed by a competition between the interaction-induced ferromagnetism and the staggered polarization of the Zeeman field across the DW. Projecting the spin-wave theory of the 2D QHFM onto the low energy 1D mode, one obtains a quadratic approximation for the dynamics in terms of an effective Luttinger liquid model \cite{Fertig_2006}. However, a field-theoretical description beyond the Gaussian level should account for the fact that, due to the vanishing of the polarizing field at the center of the DW, a semiclassical spin-wave approximation is not well-justified.

As described in the previous sections, in this paper we suggest an alternative prescription for the derivation of an effective field theory which does not fully rely on a Gaussian spin-wave approximation. We have demonstrate the possible consequences of this prescription by studying a toy model, where the full quantum dynamics of the central region of the DW is modeled by a quasi 1D spin-$1/2$ system, coupled to the spin wave fluctuations of the remaining (almost polarized) 2D ferromagnetic environment. Generally, the resulting effective 1D field theory obtained by integrating over the environmental degrees of freedom [encoded by a correction to the action Eqs. (\ref{deltaSeff}), (\ref{Veff})] is quite complex, being non-local in both space and imaginary time. This is a manifestation of the fact that the effective action can not be derived from a pure Hamiltonian dynamics of the spin system. It is interesting to note that in the long wave-length limit, the effective 1D theory describing the boundary layer of the environment [see Eqs. (\ref{G_env1D}) through (\ref{alpha_def})] can be formally mapped to a ``standard" action of a spin system with renormalized (non-quantized) spin magnitude, $s_0\rightarrow \tilde{s_0}$. Assuming $|J|s_0<B$ (which guarantees the validity of the spin-wave approximation in the 2D environment), it is implied that the spin magnitude is enlarged ($\tilde{s_0}>s_0$). This can be interpreted as an effective suppression of $\hbar$: indeed, the coupling to a polarized QHFM environment stiffens the quantum fluctuations in the DW, turning its dynamics to a more ``classical" one.

Despite the general complexity of the above mentioned effective field-theory, the presence of a high energy scale $\tilde{B}$ characterizing the gap for spin fluctuations in the environment [cf. Eq. (\ref{Veff})] implies that $\delta S_{eff}$ has a very good local approximation. This allows a mapping of the low-energy dynamics of the DW in its final form to a standard spin-$1/2$ ladder model, with renormalized parameters dictated by the coupling to the environment. In particular, the effective Zeeman field on each side of the DW center is enhanced due to the local field imposed by the environment, and anisotropy is introduced due to an effective enhancement of the exchange coupling in the $xy$-plane. We finally arrive at a sine-Gordon model [Eq. (\ref{Hamiltonian}), or equivalently Eq. (\ref{AFM})] which reflects the competition between these two effects. This reflects the possibility to obtain a quantum phase transition from Luttinger liquid behavior (encoded by the quadratic approximation) to an ordered phase with a spin gap. Recalling that the operator $S_z$ can be traced back to electric current fluctuations \cite{Abanin_2006,Shimshoni2009}, the physical interpretation of this ordered state is that of a perfect conductor: right-moving and left-moving channels (propagating along the DW) are localized in the transverse direction at opposite sides of the DW center, and backscattering is inhibited.
A possible relation of this physics to transport in single layer graphene in
the QH regime has been discussed elsewhere \cite{Shimshoni2009}.

\begin{acknowledgments}
We acknowledge useful discussions with A. Auerbach, C.-W. Huang and D. Podolsky.
This work was partially supported by the US-Israel Binational Science Foundation
(BSF) through Grant No. 2008256 , and by the US National Science Foundation
(NSF) through Grant No. DMR1005035. H. A. F. and E. S. are grateful to the hospitality of the Aspen Center for Physics (NSF 1066293), where part of this work was carried out.
\end{acknowledgments}
%______________________________________________________________________________________________________

\appendix

\section{Derivation of the effective 1D environmental Green's function}\label{appendixA}

To evaluate $\delta S_{env}^{1D}[\vec{\sigma}_0]$ (Eq. (\ref{PartFunction})), we first perform the integration over the 2D spin fields $\vec{\sigma}$
\begin{equation}
Z[\vec{\lambda}]\equiv \int D\vec{\sigma}\,
e^{i\sum\limits_{{k}_y,\omega_n}\vec{\lambda}^T({-k}_y,-\omega_n)\vec{\sigma}({k}_y,\omega_n)
-S_{2D}[\vec{\sigma} ]}\; ,
\end{equation}
noting that
\begin{equation}
\vec{\sigma}(k_y,\omega_n)=\frac{1}{\sqrt{N_x}}\sum_{k_x} \vec{\sigma}(\vec{k},\omega_n)\; .
\end{equation}
Using Eq. (\ref{S_2D}) for $S_{2D}[\vec{\sigma}]$, the Gaussian integration yields
\begin{equation}\label{Z_lambda}
Z[\vec{\lambda}]=e^{-\frac{s_0}{2}\sum\limits_{{k_y},\omega_n}\vec{\lambda}({-k_y},-\omega_n)^T
   G^{1D}({k_y},\omega_n)\vec{\lambda}({k_y},\omega_n)}
\end{equation}
in which
\begin{equation}
G^{1D}({k_y},\omega_n)\equiv \frac{1}{N_x}\sum_{k_x} G_{2D}(\vec{k},\omega_n)\; ,
\end{equation}
and $G_{2D}(\vec{k},\omega_n)$ is obtained by inverting Eq. (\ref{G_2Dinv}).
The resulting diagonal elements of $G^{1D}$ are therefore given by
\begin{eqnarray}
& &G^{1D}_{1,1}=G^{1D}_{2,2}=a\int \frac{dk_x}{2\pi}\,
 \frac{\omega_{2D}(\vec{k})}{\omega^2_{2D}(\vec{k})+\omega_n^2} \\ \nonumber
& & \cong\frac{a}{2}\int \frac{dk_x}{2\pi}\,\Big\{
 \frac{1}{|J|s_0k^2+B+i\omega_n} +\frac{1}{|J|s_0k^2+B-i\omega_n}\Big\}\\ \nonumber
 & & =\frac{a}{2}\frac{1}{|J|s_0}\int \frac{dk_x}{2\pi}\,\Big\{
 \frac{1}{k_x^2-k_-^2}+\frac{1}{k_x^2-k_+^2}\Big\}\; ,
\end{eqnarray}
where $\displaystyle k_\pm=i\sqrt{k_y^2+\frac{1}{|J|s_0}(B\pm i\omega_n)}$.
After integration, we get
\begin{eqnarray}\label{G_1D_d}
& &G^{1D}_{1,1}=G^{1D}_{2,2}  =\frac{ia}{4}\frac{1}{|J|s_0}\Big\{
 \frac{1}{k_+}+\frac{1}{k_-}\Big\}\nonumber \\
& & = \frac{ia}{4}\frac{1}{|J|s_0}\frac{k_++k_-}{k_+k_-}\; .
\end{eqnarray}
%where $\displaystyle k_x^+k_x^-=\sqrt{\Big(k_y^2+\frac{B_z^{2D}}{|J|S_0}\Big)^2+\frac{\omega_n^2}{|J|^2S_0^2}}$.
Similarly, the off-diagonal components are given by
\begin{eqnarray}\label{G_1D_od}
& &G^{1D}_{1,2}=-G^{1D}_{2,1}=a\int \frac{dk_x}{2\pi}\,
 \frac{\omega_n}{\omega^2_{2D}(\vec{k})+\omega_n^2} \nonumber \\
& &  \cong\frac{a}{4}\frac{1}{|J|s_0}\frac{k_+-k_-}{k_+k_-}\; .
\end{eqnarray}
Inserting $Z[\vec{\lambda}]$ from Eq. (\ref{Z_lambda}) (with $G^{1D}$ given by Eqs. (\ref{G_1D_d}), (\ref{G_1D_od})) into Eq. (\ref{PartFunction}), and integrating over $\vec{\lambda}$, we arrive at the final expression for the inverse Green's function $(G_{env}^{1D}({k_y},\omega_n))^{-1}$, Eq.(\ref{G_env1D}).

\bigskip

\section{Effective Action of the 1D spin-chain}\label{appendixB}

In this Appendix, we detail the final stage of derivation of the correction to the effective action, $\delta S_{eff}[\vec{S}]$ (Eq. (\ref{deltaSeff})), resulting from the interaction of the 1D spin-chain with the environment. Starting from Eq. (\ref{S_eff}), we first define the normalized complex field variables
\begin{equation}
    \bar{\varphi}=\frac{1}{\sqrt{2\tilde{s_0}}}\sigma_0^- \qquad \qquad  \varphi=\frac{1}{\sqrt{2\tilde{s_0}}}\sigma_0^+
\end{equation}
describing the environmental spins $\vec{\sigma_0}$ within a spin-wave approximation. The integral over $\vec{\sigma_0}$ can therefore be written as
\begin{widetext}
\begin{equation}
  e^{-\delta S_{eff}[\vec{S}]}=\int D\varphi D\bar{\varphi}\, \text{exp}\Big\{-S_{int}[\vec{S},\varphi,\bar{\varphi}]
-\sum\limits_{{k_y},\omega_n}
\bar{\varphi}(-{k_y},-\omega_n)
(\tilde{B}+|\tilde{J}|\tilde{s_0}k_y^2-i\omega_n)\varphi({k_y},\omega_n)\Big\}
\end{equation}
where, using Eq. (\ref{H_int}),
\begin{equation}
S_{int}[\vec{S},\varphi,\bar{\varphi}]=J'\sqrt{\frac{\tilde{s_0}}{2N_y\beta}}\sum\limits_{k_y,\omega_n}\sum_{j}\int d\tau \, \big(\varphi(k_y,\omega_n)e^{ik_yj-i\omega_n\tau}\bar{\mathcal{S}}(j,\tau) +e^{-ik_yj+i\omega_n\tau}\mathcal{S}(j,\tau)\bar{\varphi}(k_y,\omega_n)\big)\; .
\end{equation}
A straightforward Gaussian integration then yields
\begin{equation}
\delta S_{eff}[\vec{S}]=-\frac{J'^2\tilde{s_0}}{8}\sum_{j,j'}\int d\tau d\tau' \, \bar{\mathcal{S}}(j,\tau)
 \frac{1}{N_y\beta}\sum\limits_{k_y,\omega_n}e^{ik_y(j-j')-i\omega_n(\tau-\tau')}
\frac{1}{\tilde{B}+|\tilde{J}|\tilde{s_0}k_y^2-i\omega_n}\mathcal{S}(j',\tau')\; .
\end{equation}
\end{widetext}
Integrating over $k_y$ and $\omega_n$, we obtain the final expression for $\delta S_{eff}[\vec{S}]$ with the effective interaction $V_{eff}$ given by Eq. (\ref{Veff}).

\bigskip

%______________________________________________________________________________________________________
%\vfill\eject

\end{document}